\documentclass[sigconf]{acmart}

\usepackage{bm}

\AtBeginDocument{%
  }

\setcopyright{acmlicensed}
\copyrightyear{2024}
\acmYear{2024}
\acmDOI{XXXXXXX.XXXXXXX}

\acmConference[KDD'24]{AIBS 2024: The First Workshop on AI Behavioral Science}{August 25--26,
  2024}{Barcelona, Spain}
\acmISBN{978-1-4503-XXXX-X/18/06}

\begin{document}

\title{Assessing AI Utility: The Random Guesser Test for Sequential Decision-Making Systems}

\author{Shun Ide}
\email{ideshunshun@gmail.com}
\affiliation{
    \institution{Harrison High School}
    \city{Harrison}
    \state{New York}
    \country{USA}
}

\author{Allison Blunt}
\email{blunta@harrisoncsd.org}
\affiliation{
  \institution{Harrison High School}
  \city{Harrison}
  \state{New York}
  \country{USA}
}

\author{Djallel Bouneffouf}
\email{Djallel.Bouneffouf@ibm.com}
\affiliation{
  \institution{IBM T. J. Watson Research Center}
  \city{Yorktown Heights}
  \state{New York}
  \country{USA}
}

\renewcommand{\shortauthors}{Ide et al.}

\begin{abstract}
We propose a general approach to quantitatively assessing the risk and vulnerability of artificial intelligence (AI) systems to biased decisions. The guiding principle of the proposed approach is that any AI algorithm must outperform a random guesser. This may appear trivial, but empirical results from a simplistic sequential decision-making scenario involving roulette games show that sophisticated AI-based approaches often underperform the random guesser by a significant margin. We highlight that modern recommender systems may exhibit a similar tendency to favor overly low-risk options. We argue that this ``random guesser test'' can serve as a useful tool for evaluating the utility of AI actions, and also points towards increasing exploration as a potential improvement to such systems.

 
\end{abstract}


\received{28 May 2024}

\maketitle

\section{Introduction}

Recent advances in AI (artificial intelligence) technology have allowed its integration into many parts of society. As AI's involvement in high-consequence tasks grows, concerns have arisen about potential \textit{AI misalignment}~\cite{dung2023current}, a concept representing the gap between AI's actual behavior and its intended outcome. Many recent works discuss how to detect and hold AI accountable for erroneous actions. Discussions on this topic can be categorized into a few major clusters: data lineage \cite{voigt2017eu,groger2021there}, addressing integrity issues in training data; AI explainability \cite{molnar2020interpretable,dwivedi2023explainable}, dealing with challenges in understanding black-box models during high-stakes decision-making; AI fairness \cite{awad2022computational,john2022reality}, focusing on the misalignment between AI models and established norms of political correctness; and adversarial robustness \cite{chen2022adversarial}, assessing the stability of predicted classification categories in the presence of additional noise.

AI-based recommender systems \cite{portugal2018use,su2009survey,zhang2019deep} are arguably among the most widely used real-world AI applications, already integral to everyday life in online shopping and movie streaming. Modern commercial recommender systems typically combine \textit{collaborative filtering} with optimized sequential decision-making methods such as \textit{multi-armed bandits} (MAB) \cite{silva2022multi}. The key design point of MAB is balancing exploitation, which conservatively follows the best-known option, and exploration, which embraces new challenges and possibilities. MAB is a framework that aims to trade off exploitation and exploration through trial and error when there is a finite set of options. The best arm inferred from historical data may not be the optimal choice when only a limited number of samples are available. In other words, MAB is a decision-making approach that mitigates potential AI misalignment through forced exploration when the number of samples is limited.

MAB has a long and rich history of research in both theory and practice, providing a sophisticated framework to analyze potential discrepancies from an optimal course of action based on a metric called the regret \cite{lattimore2020bandit, bouneffouf2020survey, zhao2022multi}. Thompson sampling \cite{russo2018tutorial}, in particular, is an algorithm that elegantly addresses the exploration-exploitation tradeoff using posterior sampling. As long as the set of decision choices meets the criteria of AI fairness, it has minimal risk of AI misalignment in the sense that it avoids getting locked into a suboptimal choice. The use of the bandit framework is widely regarded as the best approach in many commercial recommender systems where sequential decision-making is made for the choices generated with an offline collaborative filtering algorithm.


Our main concern is that recommender systems may misalign with the actual preferences of the user, which might be perceived through annoying and stubborn internet banner advertisement on the web browser. In this work, we conduct a model study to assess potential AI misalignment in \textit{sequential decision-making} tasks such as online product recommendation. We mainly conclude that even with a sophisticated bandit algorithm, significant value misalignment can occur against common business expectations. Our study began with the assumption that Thompson sampling would be a flawless sequential decision-making approach with minimal risk of AI misalignment. Surprisingly, our experimental results show that the MAB algorithm can underperform a simple \textit{random guesser} by a significant margin in sequential roulette games, which serves as a simplest but illustrative model of online recommender systems. 

Our study has the potential to profoundly impact the detection of potential AI misalignment in real-world recommender systems, where decisions are often based on MABs. It can be used, for example, in testing a black-box personal restaurant recommendation system by providing a historical record according to a simple probabilistic model discussed later. To the best of our knowledge, this is the first work that highlights the risk of AI misalignment in MAB-based sequential decision-making settings.

\section{Related Works}

AI misalignment is a generic term that represents a discrepancy between the expected outcome from end-users and AI's actual behavior. Various forms of AI misalignment are conveniently summarized in \cite{dung2023current}. These include data lineage \cite{voigt2017eu, groger2021there}, explainability of AI (XAI) \cite{molnar2020interpretable, dwivedi2023explainable}, AI fairness \cite{awad2022computational, john2022reality}, and adversarial robustness \cite{chen2022adversarial}. Most of these topics have been motivated by concerns about the black-box nature of modern AI systems.

XAI can be viewed as a framework that detects and mitigates potential AI misalignment in a human-in-the-loop system by providing human end-users with understandable explanations of AI's decisions. Typically, XAI methods assume a batch setting \cite{molnar2020interpretable}, where training and test samples are treated independently rather than sequentially. Quantitative AI fairness analysis \cite{bellamy2019ai, lyons2021conceptualising, john2022reality} and adversarial robustness \cite{chen2022adversarial} are two significant recent advancements in the field. Although they have undergone extensive study, fairness metrics, such as disparate impact, heavily rely on specific political correctness criteria and the sensitive attributes linked to them, which may limit their general applicability to sequential recommender systems. On the other hand, adversarial robustness typically assesses the stability of predicted classification categories in the presence of additional noise, usually under a batch setting.

Sequential decision-making tasks among a given set of discrete options have commonly been formalized as a reinforcement learning (RL) problem or its simplified version called the multi-armed bandit (MAB). Due to the simplicity of choosing one of the arms (decision options) at each decision round and the availability of a highly sophisticated theoretical framework for regret analysis \cite{lattimore2020bandit, zhao2022multi}, MABs have not been considered the main subject of AI misalignment research.

Recently, Arnold et al.~ \cite{arnold2017value} discussed scenarios where a mismatch in the intended goal and actual AI outcomes exists in the context of the Markov decision process (MDP). While MDP is a general framework upon which RL and MAB models are built, their work is a high-level position paper and does not provide concrete models and empirical results. Additionally, Zhuang et al.~ \cite{zhuang2020consequences} discussed the possibility that a slight gap between an AI agent and end-users (``principal'') in the utility function could lead to potentially devastating AI misalignment. Our work may be viewed as an actual realization of their general mathematical framework. Finally, in a slightly different context, Bouneffouf et al.~\cite{bouneffouf2017bandit} attempted to characterize delusional gamblers using RL and MAB in sequential gambling tasks, including the Iowa Gambling Task, which inspired our roulette games experiments.

\section{Problem Setting }\label{sec:setting}

As mentioned previously, we focus on situations where an AI agent chooses one of the discrete actions based on historical data. Specifically, we assume that, at a time point $t$, the AI agent picks a value~$x_t$ from a finite set of actions $\mathcal{A}$, which are bet types in our case, and receives a reward $y_t$. The choice of the action is based on the historical data $\mathcal{D}^t$:
\begin{align}
\mathcal{D}^t \triangleq (x_1,y_1 ),(x_2,y_2 ),\ldots,(x_{t-1},y_{t-1}).
\end{align}
The goal of the AI agent is to maximize the total cumulative reward. The main challenge for the agent is that, especially in the beginning, the number of samples will be so limited that a ``data-driven'' decision may not be accurate. Therefore, the agent must explore options that may not necessarily considered to be the best according to~$\mathcal{D}^t$. This is the classical exploration-exploitation trade-off. 

As discussed in the Introduction, our motivation is to assess AI models' true utility. In the present context, \textit{our goal is to quantify how AI-based sequential decision-making models perform better than a simple \textbf{random guesser}}. In some sense, this is to benchmark the AI system using a random decision maker as the baseline.

\subsection{Sequential Gambling}


We employ European roulette games as a simple yet non-trivial model of recommender systems. As discussed before, commercial recommender systems typically take a two-step approach: \textit{offline} collaborative filtering to generate a relatively small number of decision choices (e.g., products to be recommended), followed by \textit{online} decision-making with MAB to choose one of the generated candidates. While typically users' demographic information is taken into account in the latter in the form of \textit{contextual bandits}, classical MABs still serve as a useful model for the online decision-making step, assuming that a specific demographic category (such as teenage boys) is focused on.



The roulette has 37 pockets from 0 through 36, to which the AI agent decides on which bet type out of $K$ choices to use with a \$1 bet. 
Table~\ref{table:bets} summarizes bet types we use, which correspond to the action space $\mathcal{A}$ with $K=3$. In this setting, $x_t$ represents one of (zero, corner, even), and $y_t$ represents the payout for a \$1 bet. In the table, $\theta_i$ and $r_i$ denote the winning probability and the payout of the $i$-th bet type, respectively. The agent loses the \$1 stake when it loses, while it receives $r_i$ in addition to the \$1 stake when winning in the $i$-th bet type. 

We use two different sets of winning probabilities. One is a fair roulette setting, where the expected values of all the bet types are the same, and a skewed roulette setting, where the zero bet has a positive expected value of \$1.  In either case, the AI agent is unaware of the true nature of the wheel. If the AI agent is intelligent enough, however, it should be able to uncover these hidden rules behind the data. In other words, an effective AI agent should be able to detect any real and exploitable patterns within the game.

\begin{table}[thb]
\caption{ Summary of bet types in European Roulette.}
\vspace{-2mm}
\begin{tabular}{ccccc}
\hline
bet type & $\theta_i$ (fair) & $\theta_i$ (skewed) & $r_i$ & expected value (fair)  \\
\hline
Zero     & 1/37    & 2/37     & 36         & $-0.027$        \\
Corner   & 4/37    & 4/37     & 8          & $-0.027$        \\
Even     & 18/37   & 18/37   & 1          & $-0.027$       \\
\hline
\end{tabular}
\label{table:bets}
\end{table}

The fair roulette setting is particularly interesting. Since the ground truth expected values are identical, the AI agent's decision strategy does not really matter in infinite betting rounds. One practical question is what would happen after a finite number of bet attempts, such as $T=50$, where $T$ denotes the number of betting rounds in a single session. This scenario is closely aligned with real-world situations involving a limited budget. This situation, where all the options have an equal expected value, has not been extensively analyzed in the standard regret analysis of MAB.

The skewed roulette setting is used to assess the impact of data non-stationalities. 
%
%
%
The importance of different stationalities in a similar scenario have also been discussed in Lin et al.~\cite{lin2019story}. Most recent works also assume that the underlying data is stationary, which makes real-life applications questionable. In our scenario, we manifest this with a situation in which the roulette wheel is skewed temporarily such that the chance to win the \textit{zero bet} doubles. 
%
Again, our agent is unaware that this change has occurred. This extra element allows us to test whether or not these agents are effectively able to explore and exploit this information.

\subsection{Algorithms Tested}

As discussed, the sequential decision-making task for a finite set of decision choices has been typically studied using reinforcement learning (RL) \cite{sutton2018reinforcement} 
in the machine learning literature. Multi-armed bandits (MAB), which address a simplified task of RL, are particularly suitable for the gambling setting \cite{Slivkins19bandits}. Among the various RL and MAB variants proposed to date, we choose the following simple algorithms that have a minimal set of model parameters:
\begin{itemize}
\item Epsilon-greedy (EG) MAB ($\varepsilon=0.1$)
\item Thompson sampling (TS) MAB
\item Temporal difference RL (TD$(\lambda)$ with $\lambda =0,1$)
\end{itemize}
Note that having less complexity in the model generally implies greater resistance to learning spurious patterns. Hence, this choice is preferable to the AI agent to suppress unwanted volatility. These models are compared with a simple \textbf{random guesser} as the baseline, which simply chooses the bet types randomly. 

In the EG MAB strategy, the AI agent computes the expected rewards for each of the action choices based on $\mathcal{D}_t$ and uses it in decision-making. Specifically, it picks the action with the highest expected reward (the ``greedy'' choice) with a probability of $1- \epsilon$, while it chooses any action randomly with a probability of $\epsilon$. 

The TS strategy learns the probability of winning for each action via Bayesian learning, which in this case involves the beta and categorical distributions as the prior and observation models, respectively. The agent learns $p_{i}^t$, the win probability of the $i$-th action at the $t$-th round, which is updated at each round with~\cite{russo2018tutorial}:
\begin{align}\label{eq:Thompson_update}
    p_{i}^t =  \frac{\alpha_i + S_i^t}{\alpha_i + \beta_i + N_i^t},
\end{align}
where $S_i^t$ is the number of wins up to the $t$-th round out of $N_i$ trials (i.e., how many times the $i$-th action has been chosen), and $\alpha_i,\beta_i$ are the parameters of the beta prior, which are initialized to ones.

Finally, in this setting, the TD strategy evaluates the expected cumulative reward (called the Q-value) for each action with no state variable defined and chooses the action with the highest expected cumulative reward. In TD(0), when an action $i$ is chosen, the Q-value is updated as 
\begin{align}
    Q_i \leftarrow Q_i + \alpha (r_i^t - Q_i),
\end{align}
where $r^t_i$ is the observed reward at the $t$-th round. The Q value is initialized to zero, and $\alpha$ is set to 0.1 in our experiments. In TD(1), the empirical average is computed with $\mathcal{D}^t$ for $Q_i$. For more details on the TD-learning, see a standard RL textbook, e.g.,  \cite{sutton2018reinforcement}.

\section{Empirical Evaluation}\label{sec:experiments}

This section presents the results of the empirical evaluation of the decision-making algorithms introduced in the previous section.\footnote{
A Python implementation is available at https://github.com/ShunIde/Python-Implementation-for-AIBS-2024-Submission
} We employed two metrics to evaluate utility. One is the \textit{rate of successful sessions} under a finite session horizon $T$, where a session (i.e., a sequence of bet rounds) is considered successful when the balance is nonnegative at the completion of the final round $t=T$. The other metric we used is the \textit{number of rounds until bankruptcy}, defined as the number of bet rounds until the balance becomes zero or below. For the latter, no finite session horizon is assumed, and hence, the metric can theoretically be positive infinite.

These metrics were statistically tested using a one-way analysis of variance (ANOVA). Specifically, it compared the results of the AI algorithms to the control group, which consists of samples generated by the random guesser.

\subsection{Success Over Rounds}



We ran roulette simulations under two session horizons: $T=50$ and $500$ rounds, both using the \textit{fair roulette}. The final balance of each session was recorded to count the number of successful sessions. For each of the four AI algorithms as well as the random baseline, we repeated the gambling simulation 10,000 times with different random seeds.

\begin{figure}[th]
    \centering
    \includegraphics[width=0.4\textwidth, height=6cm]{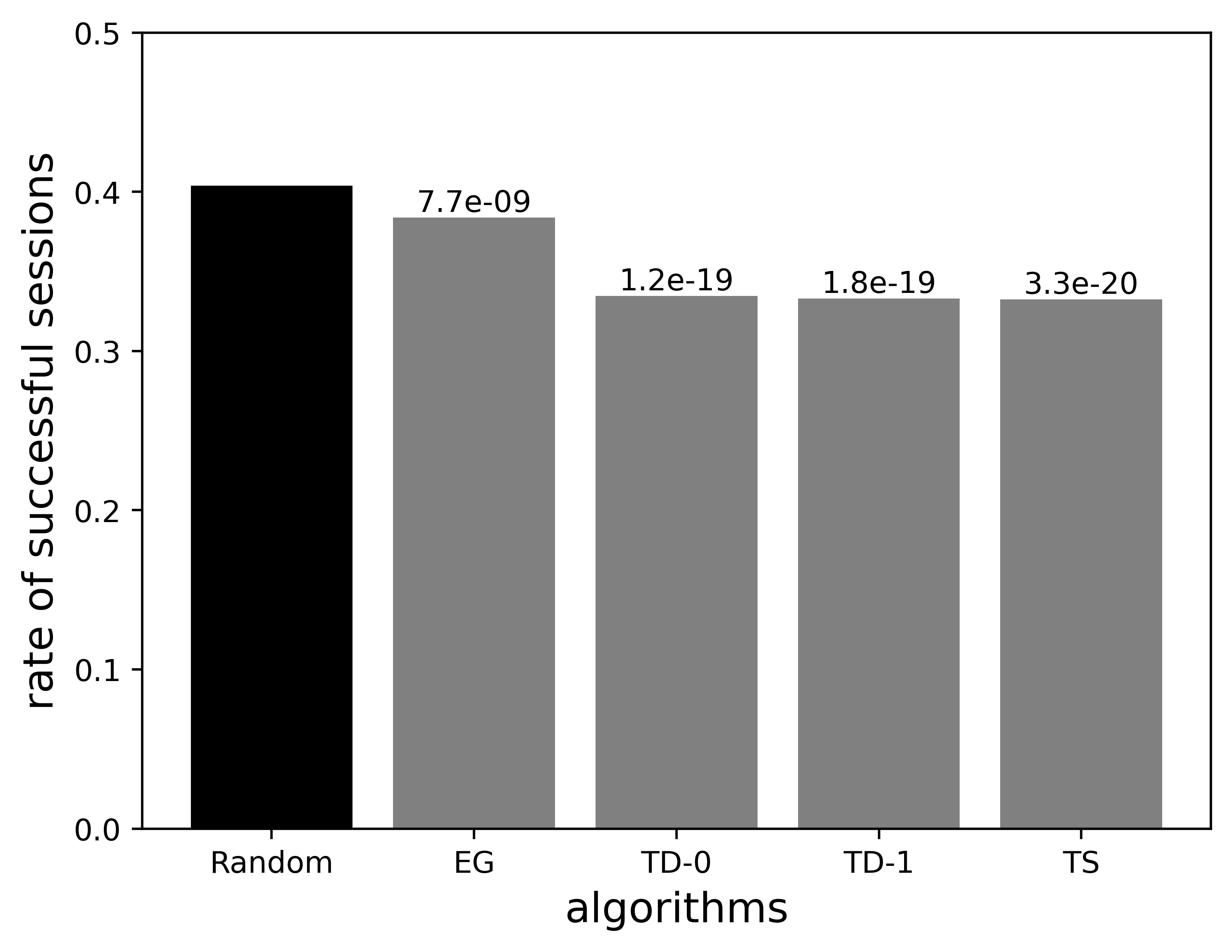}
    \caption{Recorded simulations with a net profit at 50 rounds. Random strategy had the most profit, at roughly a 40\% rate.}
    \label{fig:histogram1}
\end{figure}

\begin{figure}[th]
    \centering
    \includegraphics[width=0.4\textwidth, height=6cm]{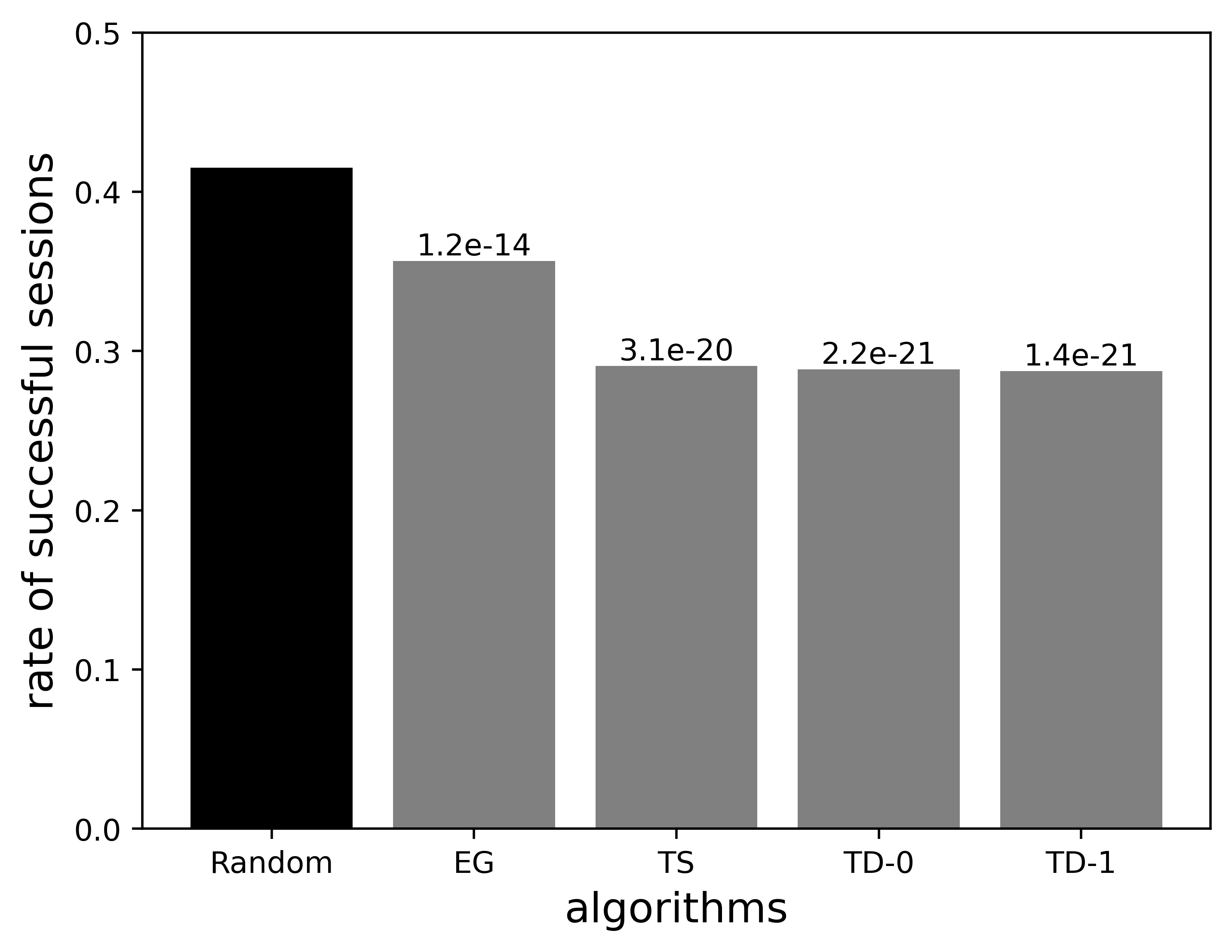}
    \caption{Recorded simulations with a net profit at 500 rounds. Larger gap develops between random and other groups.}
    \label{fig:histogram2}
\end{figure}

Figures~\ref{fig:histogram1} and \ref{fig:histogram2} show the success rates for $T=50$ and $500$, respectively. The $p$-values of the one-way ANOVA are shown at the top of the bars: The smaller the $p$-value is, the more statistically significant the difference is. Surprisingly, \textit{the random guesser performed significantly better than the AI agents}. Among the four AI algorithms, the more sophisticated algorithms (TS and TD$(\lambda)$) performed worse than the simpler one (EG). The trend is consistent between $T=50$ and $500$. 

The results appear counter-intuitive given that the three bet types have the same expected value in the fair roulette. This outcome can be explained by the preference for high-risk and high-return options of the random guesser and the relatively large volatility of the balance under a finite $T$. In fact, in Thompson sampling, one can mathematically evaluate the probability that the $k$-th action out of $K$ choices provides the highest reward as 
\begin{align} \label{eq:max_prob}    P_k = \frac{\theta_k}{1-\prod_{j=1}^K(1-\theta_j)}\prod_{i:\  r_i > r_k}(1-\theta_i).
\end{align}
We omit the derivation due to space limitations. In our setting, we have $(P_1,P_2,P_3)\approx(0.05, 0.19, 0.76)$ for the zero, corner, and even bets, respectively. This means that in those AI algorithms, the zero bet option is rarely chosen, which contrasts sharply with the random guesser who chooses the zero bet with a 1/3 chance. Under a finite $T$, this risk-taking tendency appears to be rewarding. 

As long as $T\leq 500$ in our setting, we conclude that the well-thought-out AI algorithms do not provide an optimal strategy. This calls for significant attention in real-world recommender systems, as there is a possibility that these systems may overly favor low-risk and low-return options.

\subsection{Stationary and Nonstationary Scenario}

In addition to comparing the success of algorithms over time, their ability to survive economically was also compared. In this scenario, the agent starts with the same setting, but the simulation is set to end when the gambler's balance hits zero. The number of rounds it took for the agent to go bankrupt was recorded. 

\begin{figure}[th]
    \centering
    \includegraphics[width=0.4\textwidth, height=6cm]{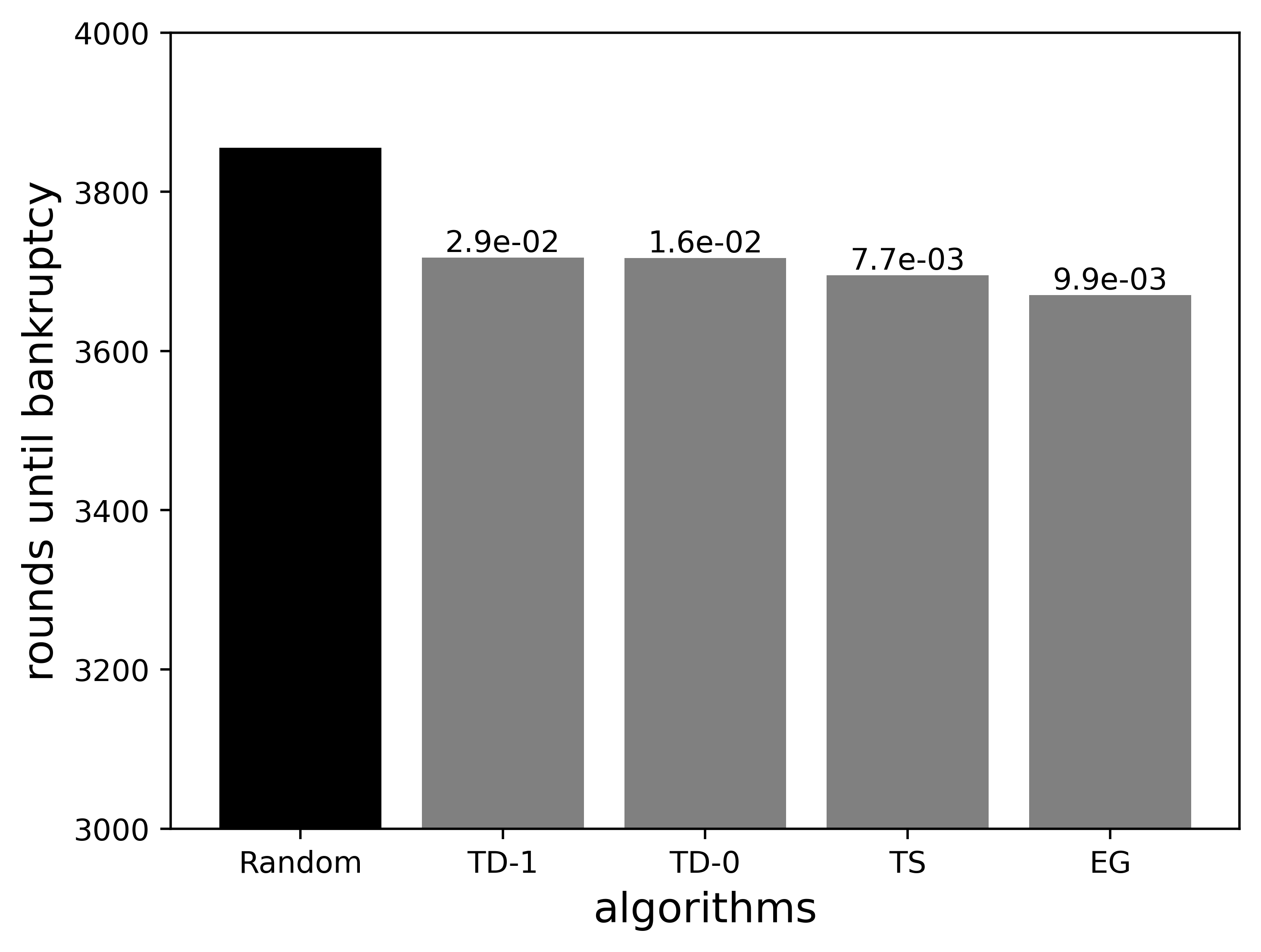}
    \caption{Survival of algorithms in stationary scenario. Random survives for significantly longer.}
    \label{fig:histogram3}
\end{figure}

For comparison purposes, Figure~\ref{fig:histogram3} plots the number of rounds until bankruptcy, or simply survival, of the algorithms in the stationary scenario as previously discussed. Again, the $p$-values from the one-way ANOVA are shown at the top of the bars. Consistent with the previous result, the random guesser lasts for longer, i.e., outperforms the sophisticated AI agents. 

\begin{figure}[th]
    \centering
    \includegraphics[width=0.4\textwidth, height=6cm]{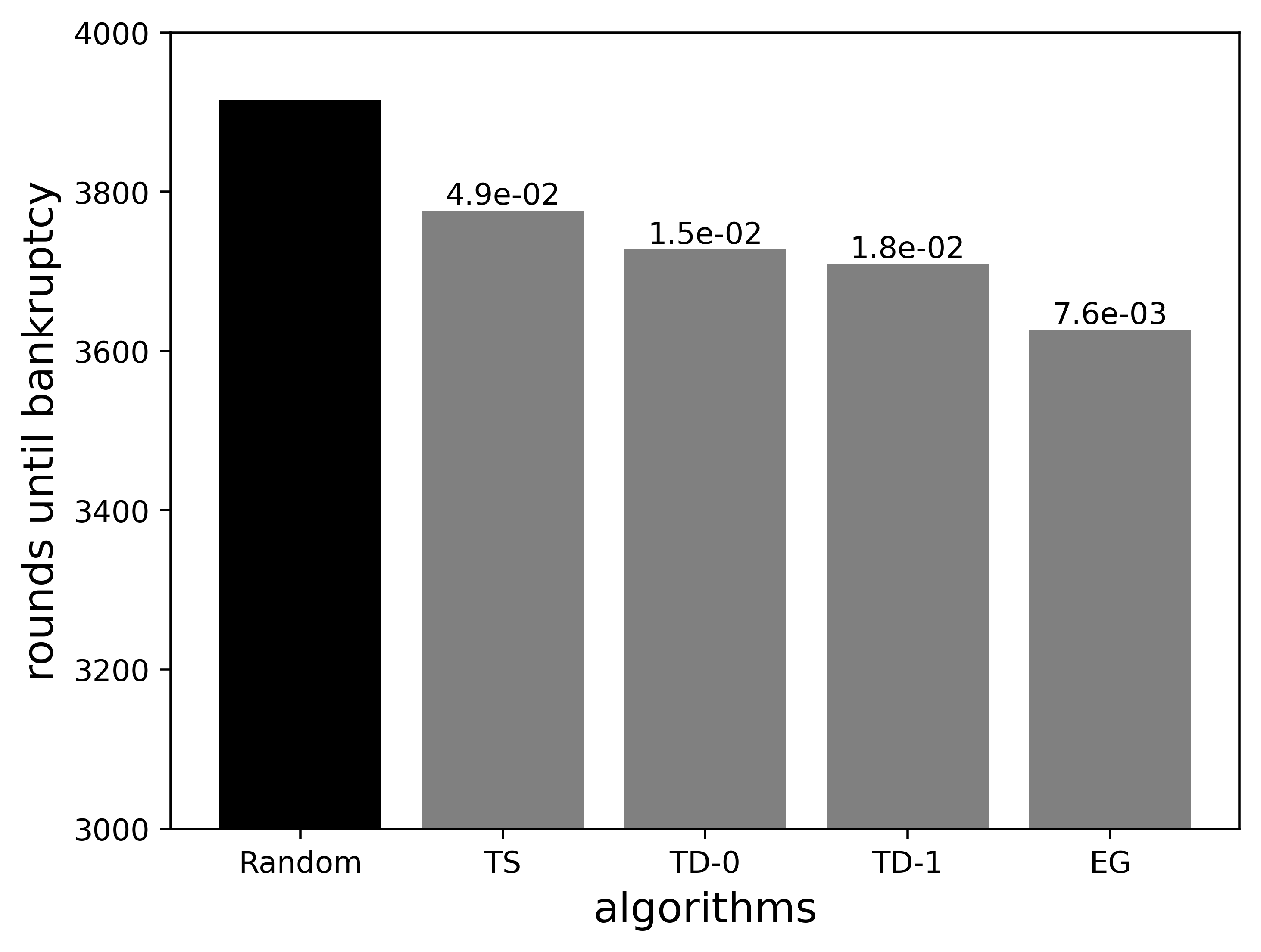}
    \caption{Survival of algorithms in nonstationary scenario. TS seems to have taken advantage of the situation, but nevertheless bankrupts faster than Random.}
    \label{fig:histogram4}
\end{figure}

Figure~\ref{fig:histogram4} explores a nonstationary scenario, where the winning probability of the zero bet increases in $100 \leq t \leq 200$. Thompson Sampling seems to have made an attempt at capitalizing on the zero bet, but is still outlived by the random group. This again shows the detrimental effects of the AI agent over-neglecting the exploration option.




\section{Concluding Remarks}

The roulette experiment serves as a simple yet illustrative example of how AI algorithms can fail the random guesser test. We have found that the sophisticated RL algorithms underperformed compared to the random guesser in a certain controlled setting. The use of a random guesser allows for the quantitative evaluation of the vulnerabilities and risks associated with AI systems in the sequential decision-making scenario. In particular, the empirical results suggest that commercial recommender systems may overly prefer low-risk and low-return options. Empirically, we showed that the AI agents were only rarely selecting the high-reward \textit{zero bet} in favor of the \textit{even bet} which occurred far more often. This could potentially be an explanation of the behavior of some annoying online advertisements. The AI overly favors safe options, neglecting other potentially valuable options. For instance, when a consumer directs themselves to a product, the AI would favor that one selection too heavily, which may not accurately suit the consumer's tastes. In summary, our results suggest that a potential solution to improving recommender systems would be to increase the exploration parameter.

For future work, we plan to extend this framework to state-of-the-art sequence prediction models. We are also interested in making further connections between the neglect of exploration and AI misalignment.

\bibliographystyle{ACM-Reference-Format}
\bibliography{overlearning}


\begin{thebibliography}{24}


\ifx \showCODEN    \undefined \def \showCODEN     #1{\unskip}     \fi
\ifx \showDOI      \undefined \def \showDOI       #1{#1}\fi
\ifx \showISBNx    \undefined \def \showISBNx     #1{\unskip}     \fi
\ifx \showISBNxiii \undefined \def \showISBNxiii  #1{\unskip}     \fi
\ifx \showISSN     \undefined \def \showISSN      #1{\unskip}     \fi
\ifx \showLCCN     \undefined \def \showLCCN      #1{\unskip}     \fi
\ifx \shownote     \undefined \def \shownote      #1{#1}          \fi
\ifx \showarticletitle \undefined \def \showarticletitle #1{#1}   \fi
\ifx \showURL      \undefined \def \showURL       {\relax}        \fi
\providecommand\bibfield[2]{#2}
\providecommand\bibinfo[2]{#2}
\providecommand\natexlab[1]{#1}
\providecommand\showeprint[2][]{arXiv:#2}

\bibitem[Arnold et~al\mbox{.}(2017)]%
        {arnold2017value}
\bibfield{author}{\bibinfo{person}{Thomas Arnold}, \bibinfo{person}{Daniel Kasenberg}, {and} \bibinfo{person}{Matthias Scheutz}.} \bibinfo{year}{2017}\natexlab{}.
\newblock \showarticletitle{Value Alignment or Misalignment--What Will Keep Systems Accountable?}. In \bibinfo{booktitle}{\emph{Workshops at the thirty-first AAAI conference on artificial intelligence}}.
\newblock


\bibitem[Awad~et al.(2022)]%
        {awad2022computational}
\bibfield{author}{\bibinfo{person}{E. Awad~et al.}} \bibinfo{year}{2022}\natexlab{}.
\newblock \showarticletitle{Computational ethics}.
\newblock \bibinfo{journal}{\emph{Trends in Cognitive Sciences}} \bibinfo{volume}{26}, \bibinfo{number}{5} (\bibinfo{year}{2022}), \bibinfo{pages}{388--405}.
\newblock


\bibitem[Bellamy~et al.(2019)]%
        {bellamy2019ai}
\bibfield{author}{\bibinfo{person}{R.K.E. Bellamy~et al.}} \bibinfo{year}{2019}\natexlab{}.
\newblock \showarticletitle{{AI} Fairness 360: An extensible toolkit for detecting and mitigating algorithmic bias}.
\newblock \bibinfo{journal}{\emph{IBM Journal of Research and Development}} \bibinfo{volume}{63}, \bibinfo{number}{4/5} (\bibinfo{year}{2019}), \bibinfo{pages}{4--1}.
\newblock


\bibitem[Bouneffouf et~al\mbox{.}(2020)]%
        {bouneffouf2020survey}
\bibfield{author}{\bibinfo{person}{Djallel Bouneffouf}, \bibinfo{person}{Irina Rish}, {and} \bibinfo{person}{Charu Aggarwal}.} \bibinfo{year}{2020}\natexlab{}.
\newblock \showarticletitle{Survey on applications of multi-armed and contextual bandits}. In \bibinfo{booktitle}{\emph{2020 IEEE Congress on Evolutionary Computation (CEC)}}. IEEE, \bibinfo{pages}{1--8}.
\newblock


\bibitem[Bouneffouf~et al.(2017)]%
        {bouneffouf2017bandit}
\bibfield{author}{\bibinfo{person}{D. Bouneffouf~et al.}} \bibinfo{year}{2017}\natexlab{}.
\newblock \showarticletitle{Bandit models of human behavior: Reward processing in mental disorders}. In \bibinfo{booktitle}{\emph{Proc. 10th Intl. Conf. Artificial General Intelligence (AGI 17)}}. Springer, \bibinfo{pages}{237--248}.
\newblock


\bibitem[Chen~et al.(2022)]%
        {chen2022adversarial}
\bibfield{author}{\bibinfo{person}{P.-Y. Chen~et al.}} \bibinfo{year}{2022}\natexlab{}.
\newblock \bibinfo{booktitle}{\emph{Adversarial robustness for machine learning}}.
\newblock \bibinfo{publisher}{Academic Press}.
\newblock


\bibitem[Dung(2023)]%
        {dung2023current}
\bibfield{author}{\bibinfo{person}{Leonard Dung}.} \bibinfo{year}{2023}\natexlab{}.
\newblock \showarticletitle{Current cases of AI misalignment and their implications for future risks}.
\newblock \bibinfo{journal}{\emph{Synthese}} \bibinfo{volume}{202}, \bibinfo{number}{5} (\bibinfo{year}{2023}), \bibinfo{pages}{138}.
\newblock


\bibitem[Dwivedi~et al.(2023)]%
        {dwivedi2023explainable}
\bibfield{author}{\bibinfo{person}{R. Dwivedi~et al.}} \bibinfo{year}{2023}\natexlab{}.
\newblock \showarticletitle{Explainable {AI} ({XAI}): Core ideas, techniques, and solutions}.
\newblock \bibinfo{journal}{\emph{Comput. Surveys}} \bibinfo{volume}{55}, \bibinfo{number}{9} (\bibinfo{year}{2023}), \bibinfo{pages}{1--33}.
\newblock


\bibitem[Gr{\"o}ger(2021)]%
        {groger2021there}
\bibfield{author}{\bibinfo{person}{C. Gr{\"o}ger}.} \bibinfo{year}{2021}\natexlab{}.
\newblock \showarticletitle{There is no {AI} without data}.
\newblock \bibinfo{journal}{\emph{Commun. ACM}} \bibinfo{volume}{64}, \bibinfo{number}{11} (\bibinfo{year}{2021}), \bibinfo{pages}{98--108}.
\newblock


\bibitem[John-Mathews~et al.(2022)]%
        {john2022reality}
\bibfield{author}{\bibinfo{person}{J.-M. John-Mathews~et al.}} \bibinfo{year}{2022}\natexlab{}.
\newblock \showarticletitle{From reality to world. A critical perspective on {AI} fairness}.
\newblock \bibinfo{journal}{\emph{Journal of Business Ethics}} \bibinfo{volume}{178}, \bibinfo{number}{4} (\bibinfo{year}{2022}), \bibinfo{pages}{945--959}.
\newblock


\bibitem[Lattimore and Szepesv{\'a}ri(2020)]%
        {lattimore2020bandit}
\bibfield{author}{\bibinfo{person}{Tor Lattimore} {and} \bibinfo{person}{Csaba Szepesv{\'a}ri}.} \bibinfo{year}{2020}\natexlab{}.
\newblock \bibinfo{booktitle}{\emph{Bandit algorithms}}.
\newblock \bibinfo{publisher}{Cambridge University Press}.
\newblock


\bibitem[Lin et~al\mbox{.}(2019)]%
        {lin2019story}
\bibfield{author}{\bibinfo{person}{Baihan Lin}, \bibinfo{person}{Guillermo Cecchi}, \bibinfo{person}{Djallel Bouneffouf}, \bibinfo{person}{Jenna Reinen}, {and} \bibinfo{person}{Irina Rish}.} \bibinfo{year}{2019}\natexlab{}.
\newblock \showarticletitle{A story of two streams: Reinforcement learning models from human behavior and neuropsychiatry}.
\newblock \bibinfo{journal}{\emph{arXiv preprint arXiv:1906.11286}} (\bibinfo{year}{2019}).
\newblock


\bibitem[Lyons~et al.(2021)]%
        {lyons2021conceptualising}
\bibfield{author}{\bibinfo{person}{H. Lyons~et al.}} \bibinfo{year}{2021}\natexlab{}.
\newblock \showarticletitle{Conceptualising contestability: Perspectives on contesting algorithmic decisions}.
\newblock \bibinfo{journal}{\emph{Proceedings of the ACM on Human-Computer Interaction}} \bibinfo{volume}{5}, \bibinfo{number}{CSCW1} (\bibinfo{year}{2021}), \bibinfo{pages}{1--25}.
\newblock


\bibitem[Molnar(2020)]%
        {molnar2020interpretable}
\bibfield{author}{\bibinfo{person}{C. Molnar}.} \bibinfo{year}{2020}\natexlab{}.
\newblock \bibinfo{booktitle}{\emph{Interpretable machine learning}}.
\newblock \bibinfo{publisher}{Lulu. com}.
\newblock


\bibitem[Portugal et~al\mbox{.}(2018)]%
        {portugal2018use}
\bibfield{author}{\bibinfo{person}{Ivens Portugal}, \bibinfo{person}{Paulo Alencar}, {and} \bibinfo{person}{Donald Cowan}.} \bibinfo{year}{2018}\natexlab{}.
\newblock \showarticletitle{The use of machine learning algorithms in recommender systems: A systematic review}.
\newblock \bibinfo{journal}{\emph{Expert Systems with Applications}}  \bibinfo{volume}{97} (\bibinfo{year}{2018}), \bibinfo{pages}{205--227}.
\newblock


\bibitem[Russo et~al\mbox{.}(2018)]%
        {russo2018tutorial}
\bibfield{author}{\bibinfo{person}{Daniel~J Russo}, \bibinfo{person}{Benjamin Van~Roy}, \bibinfo{person}{Abbas Kazerouni}, \bibinfo{person}{Ian Osband}, \bibinfo{person}{Zheng Wen}, {et~al\mbox{.}}} \bibinfo{year}{2018}\natexlab{}.
\newblock \showarticletitle{A tutorial on thompson sampling}.
\newblock \bibinfo{journal}{\emph{Foundations and Trends{\textregistered} in Machine Learning}} \bibinfo{volume}{11}, \bibinfo{number}{1} (\bibinfo{year}{2018}), \bibinfo{pages}{1--96}.
\newblock


\bibitem[Silva et~al\mbox{.}(2022)]%
        {silva2022multi}
\bibfield{author}{\bibinfo{person}{N{\'\i}collas Silva}, \bibinfo{person}{Heitor Werneck}, \bibinfo{person}{Thiago Silva}, \bibinfo{person}{Adriano~CM Pereira}, {and} \bibinfo{person}{Leonardo Rocha}.} \bibinfo{year}{2022}\natexlab{}.
\newblock \showarticletitle{Multi-armed bandits in recommendation systems: A survey of the state-of-the-art and future directions}.
\newblock \bibinfo{journal}{\emph{Expert Systems with Applications}}  \bibinfo{volume}{197} (\bibinfo{year}{2022}), \bibinfo{pages}{116669}.
\newblock


\bibitem[Slivkins(2019)]%
        {Slivkins19bandits}
\bibfield{author}{\bibinfo{person}{Aleksandrs Slivkins}.} \bibinfo{year}{2019}\natexlab{}.
\newblock \showarticletitle{Introduction to Multi-Armed Bandits}.
\newblock \bibinfo{journal}{\emph{Foundations and Trends{\textregistered} in Machine Learning}} \bibinfo{volume}{12}, \bibinfo{number}{1-2} (\bibinfo{year}{2019}), \bibinfo{pages}{1--286}.
\newblock


\bibitem[Su and Khoshgoftaar(2009)]%
        {su2009survey}
\bibfield{author}{\bibinfo{person}{Xiaoyuan Su} {and} \bibinfo{person}{Taghi~M Khoshgoftaar}.} \bibinfo{year}{2009}\natexlab{}.
\newblock \showarticletitle{A survey of collaborative filtering techniques}.
\newblock \bibinfo{journal}{\emph{Advances in artificial intelligence}}  \bibinfo{volume}{2009} (\bibinfo{year}{2009}).
\newblock


\bibitem[Sutton and Barto(2018)]%
        {sutton2018reinforcement}
\bibfield{author}{\bibinfo{person}{Richard~S Sutton} {and} \bibinfo{person}{Andrew~G Barto}.} \bibinfo{year}{2018}\natexlab{}.
\newblock \bibinfo{booktitle}{\emph{Reinforcement learning: An introduction}}.
\newblock \bibinfo{publisher}{MIT press}.
\newblock


\bibitem[Voigt~et al.(2017)]%
        {voigt2017eu}
\bibfield{author}{\bibinfo{person}{P. Voigt~et al.}} \bibinfo{year}{2017}\natexlab{}.
\newblock \showarticletitle{The {EU} general data protection regulation ({GDPR})}.
\newblock \bibinfo{journal}{\emph{A Practical Guide, 1st Ed.}} \bibinfo{volume}{10}, \bibinfo{number}{3152676} (\bibinfo{year}{2017}), \bibinfo{pages}{10--5555}.
\newblock


\bibitem[Zhang et~al\mbox{.}(2019)]%
        {zhang2019deep}
\bibfield{author}{\bibinfo{person}{Shuai Zhang}, \bibinfo{person}{Lina Yao}, \bibinfo{person}{Aixin Sun}, {and} \bibinfo{person}{Yi Tay}.} \bibinfo{year}{2019}\natexlab{}.
\newblock \showarticletitle{Deep learning based recommender system: A survey and new perspectives}.
\newblock \bibinfo{journal}{\emph{ACM computing surveys (CSUR)}} \bibinfo{volume}{52}, \bibinfo{number}{1} (\bibinfo{year}{2019}), \bibinfo{pages}{1--38}.
\newblock


\bibitem[Zhao(2022)]%
        {zhao2022multi}
\bibfield{author}{\bibinfo{person}{Qing Zhao}.} \bibinfo{year}{2022}\natexlab{}.
\newblock \bibinfo{booktitle}{\emph{Multi-armed bandits: Theory and applications to online learning in networks}}.
\newblock \bibinfo{publisher}{Springer Nature}.
\newblock


\bibitem[Zhuang and Hadfield-Menell(2020)]%
        {zhuang2020consequences}
\bibfield{author}{\bibinfo{person}{Simon Zhuang} {and} \bibinfo{person}{Dylan Hadfield-Menell}.} \bibinfo{year}{2020}\natexlab{}.
\newblock \showarticletitle{Consequences of misaligned AI}.
\newblock \bibinfo{journal}{\emph{Advances in Neural Information Processing Systems}}  \bibinfo{volume}{33} (\bibinfo{year}{2020}), \bibinfo{pages}{15763--15773}.
\newblock


\end{thebibliography}

\appendix

\end{document}